\documentclass[fleqn,10pt]{wlscirep}
\usepackage{placeins} 
\usepackage{float}
\usepackage{subcaption} 
\usepackage[utf8]{inputenc}
\usepackage{natbib}
\usepackage{pdflscape}
\usepackage{graphicx}
\usepackage{caption}
\usepackage[left]{lineno}

\usepackage[T1]{fontenc}
\usepackage[printonlyused,withpage,nolist]{acronym}
\usepackage[font=small]{caption} 

\title{ Seismic Depth Imaging of the 2024 Noto Earthquake (M7.6) Rupture Area}

\author[1*]{Hamzeh Mohammadigheymasi}
\author[1]{Jin-Oh Park}

\affil[1]{Atmosphere and Ocean Research Institute, The University of Tokyo, 277-0882, Kashiwa, Japan}
\affil[*]{hamzeh.prof@gmail.com}

\begin{abstract}
On January 1, 2024, at 07:10:22.5 UTC, a moment magnitude ($M_w$) 7.6 earthquake struck the Noto Peninsula, Japan, causing intense ground shaking and triggering a tsunami along the eastern coast of the Japan Sea. Preliminary analysis by the Japan Meteorological Agency (JMA) identified the event as a reverse-fault rupture, consistent with a northwest–southeast-oriented compressional stress regime. Subsequent aftershock distribution analysis (JMA, 2024) revealed that the causative fault extended approximately 150 km, from the western coast of the Noto Peninsula to the northeastern offshore area, aligning with the inferred tsunami source region. While extensive studies have been conducted on the rupture mechanism and seismic impacts, high-resolution seismic imaging of the shallow crustal structure within the rupture zone remains limited. To address this gap, the Atmosphere and Ocean Research Institute (AORI) at the University of Tokyo conducted a multichannel seismic (MCS) reflection survey aboard the R/V Hakuho-Maru between March 4 and 16, 2024. The survey collected high-quality MCS data along 14 seismic profiles, each approximately 45 km in length. The acquired data were processed using an advanced depth imaging workflow that incorporated grid-based tomography refined by automated continuity attributes to enhance reflection coherency. Structural attributes such as dip and continuity were extracted from the migrated sections and used for automated horizon picking via seismic pencil construction. The P-wave velocity model was iteratively refined using grid-based tomography to optimize horizon alignment and minimize residual moveout (RMO) in the migrated common image gathers. The resulting 2D seismic sections (and their 3D visualizations), presented in this report provide the first high-resolution images of the shallow rupture zone associated with the 2024 Noto earthquake. This dataset offers a critical foundation for ongoing research into fault geometry, rupture dynamics, and the broader seismotectonic framework of the region.

\end{abstract}
\begin{document}

\flushbottom
\maketitle
%
%
\thispagestyle{empty}

\section*{}

The Sea of Japan, bordered by the Japanese archipelago, the Korean Peninsula, and Russia, exhibits a complex tectonic environment shaped by back-arc extension followed by subsequent compression. The opening of the Sea of Japan occurred during the late Oligocene to middle Miocene (approximately 28–15 Ma), driven by extensional tectonics associated with the rollback of the Pacific Plate subduction along the western Pacific margin \cite{Jolivet1994, Lallemand2009}. This extensional phase led to rifting along the eastern Eurasian Plate margin, forming oceanic basins and widespread north-south normal faulting that defined the region's structural framework \citep{Kimura1996, Takeuchi2002, Lallemand2009}. By the mid-Miocene, seafloor spreading ceased, initiating tectonic inversion and a transition to east-west compression by the late Pliocene (\textasciitilde
3–5 Ma) \citep{Tamaki1985}. This compressional regime, driven by the ongoing subduction of the Pacific Plate beneath Japan, reactivated pre-existing extensional faults as reverse or thrust faults, resulting in the formation of fold-and-thrust belts along the eastern margin of the Sea of Japan \citep{Sato2020}.

\begin{figure*}[ht!]
	\centering
	\includegraphics[width=0.6\textwidth]{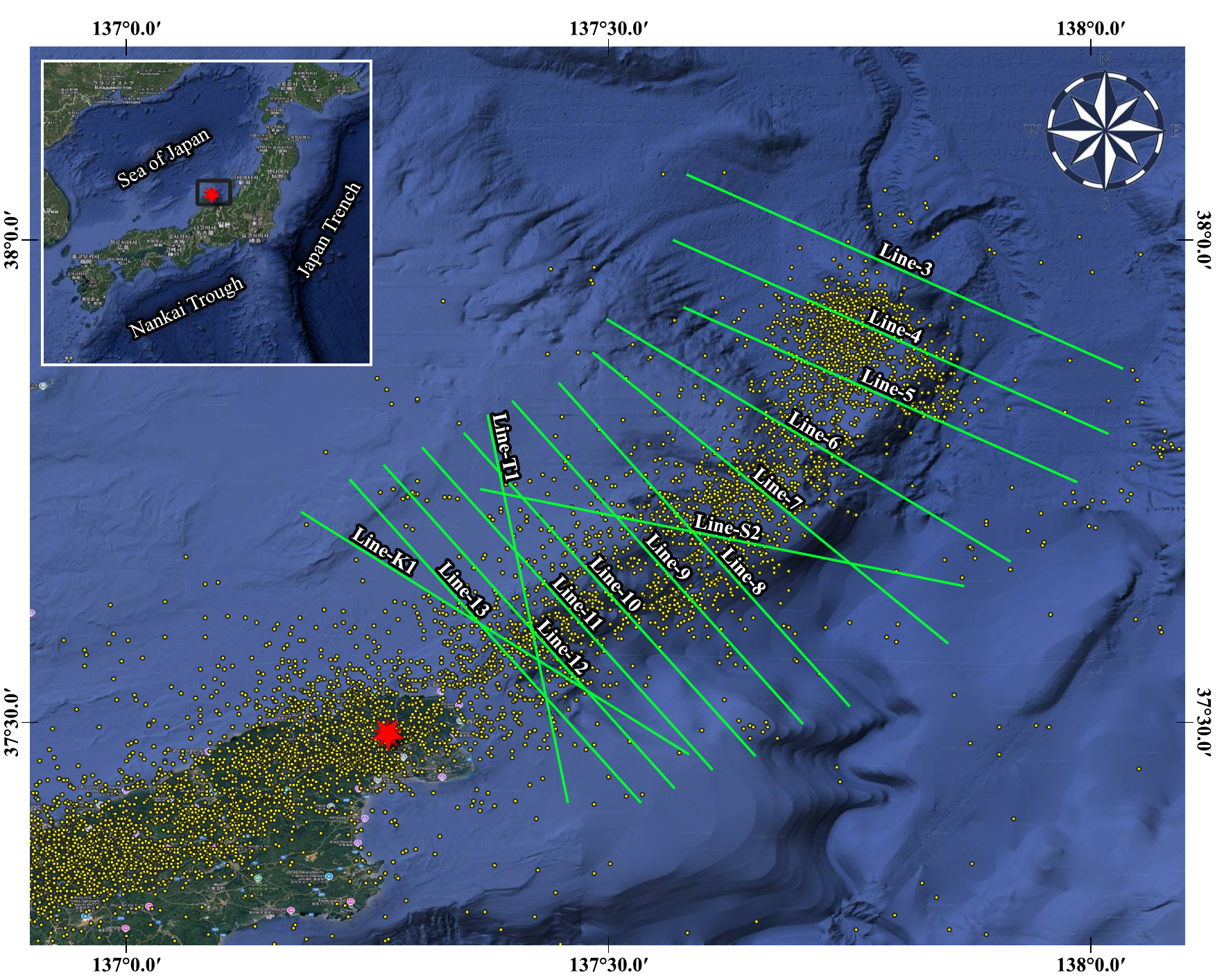}
	\caption{The inset map provides a regional overview of Japan and its surrounding areas, with the epicenter of the 2024 Noto earthquake marked by a red star. A black rectangle outlines the area covered in the main figure. The main panel displays the trajectories of the \ac{2D} seismic lines acquired during the NOTO 2024 Multichannel Seismic (MCS) survey conducted in March 2024. Seismic profiles are shown as light green lines. Yellow circles denote aftershocks recorded between January 1 and January 15, 2024 (JMA), distributed along the approximately 150 km-long active fault zone associated with the mainshock.}
	\label{Area}
\end{figure*}

Seismicity along the eastern margin of the Sea of Japan reflects the ongoing tectonic inversion of this back-arc basin, where compressional reactivation of inherited extensional structures has generated a series of destructive earthquakes. Although such events are relatively infrequent, as evidenced by the 1940 Shakotan ($M_w \approx 7.5$), 1964 Niigata ($M_w = 7.5$–7.6), 1983 Nihonkai-Chūbu ($M_w = 7.8$), and 1993 Hokkaido Nansei-Oki ($M_w = 7.7$) earthquakes \citep{Watanabe2000}, they pose a significant seismic hazard to the surrounding coastal regions. The January 1, 2024, $M_w = 7.6$ reverse-fault earthquake that struck the Noto Peninsula is the most recent and impactful example, marking one of the largest seismic events in the Sea of Japan region in recent decades (see inset in Fig.~\ref{Area}). While the earthquake caused intense ground shaking and generated a significant tsunami, it also presents a rare opportunity to investigate active crustal deformation and rupture dynamics within an inverted back-arc setting.

A more detailed understanding of the rupture geometry and fault behavior associated with the 2024 Noto earthquake is essential, particularly in light of the significant tsunami that accompanied the event. The rupture area extended northeastward, offshore of the peninsula, coinciding with the region where the tsunami was most strongly generated. This suggests that displacement along a deep seismogenic fault likely produced shallow crustal movement near the seafloor, which in turn contributed to tsunami generation. The source fault, inferred from aftershock distributions, stretches approximately 150 km from the west coast of the Noto Peninsula to the offshore area northeast of the peninsula. A portion of this fault, which experienced a maximum slip of about 4 meters, is considered responsible for the tsunami in the northeastern offshore region of the Noto Peninsula \citep{fujii2024slip, Kutschera2024}. However, significant uncertainties remain regarding the structure, mechanical properties, and rupture dynamics of this tsunamigenic fault, particularly concerning how shallow crustal deformation and fluid pressure affect fault slip behavior.

Numerous studies have investigated the underlying causes and rupture mechanisms of the 2024 Noto Peninsula earthquake. By relocating earthquake hypocenters, \cite{yoshida2024role} identified the critical role of a previously unrecognized fault in the early rupture process. Their study revealed that foreshock interactions, aseismic slip, and fluid migration along this hidden fault facilitated the initiation and propagation of the main shock. Through spectral and wavelet analysis of Toyama Bay tides and wave gauge records, \cite{mulia2024compounding} demonstrated that short-period tsunami waves were likely triggered by a submarine landslide. Their combined earthquake-landslide source model successfully reproduced observed tsunami waveforms, emphasizing the contribution of the landslide to tsunami generation. 
Using a similar approach, \cite{fujii2024slip} analyzed tsunami waveforms from six wave gauges and 12 tide gauges around the Sea of Japan, along with GNSS data from 53 stations on the Noto Peninsula, to estimate slip distribution and seismic moments on active faults. Their results, interpreted based on the \ac{JSPJ} model, indicate that coseismic slips of 3.5, 3.2, and 3.2 m occurred on sub-faults NT4, NT5, and NT6, which are located on the northern coast of the Noto Peninsula and dip southeastward. Additionally, a smaller slip of 1.0 m, estimated on NT8 at the southwestern end of the 2024 rupture, may be attributed to its previous rupture during the 2007 Noto earthquake. They also estimated that the total length of these four faults is approximately 100 km.

Through near-source waveform analysis and source imaging techniques, \citet{xu2024dual} demonstrated that the 2024 $M_w$ 7.6 Noto Peninsula earthquake was initiated by the failure of a high-stress drop asperity. This asperity initially impeded rupture propagation but eventually failed due to the interaction of bilateral rupture fronts, transforming the event into a large-scale seismic rupture. Their study further indicated that this asperity remained intact during earlier seismic swarms, acting as a temporary barrier that suppressed full rupture development. Its eventual failure, marked by a concentrated release of stress, was central to the earthquake’s escalation, underscoring the critical role of localized fault properties and rupture complexity in swarm-triggered seismicity.

Despite growing interest in the rupture mechanics of this event, detailed knowledge of the shallow crustal structure in the source region remains limited. This is largely due to the lack of high-resolution seismic imaging, which is essential for delineating fault geometry, near-surface deformation, and potential tsunamigenic features. Reliable depth-domain seismic imaging requires high-fidelity \ac{MCS} reflection data and the implementation of advanced depth imaging workflows. At the core of such workflows is the estimation of a precise subsurface velocity model, typically achieved using grid-based tomography, a technique widely adopted in the seismic exploration industry \citep{ebigbo2016influence, kosloff2008velocity, liao2009seismic, woodward2008decade, bradford2006imaging}.

To address this limitation, the \ac{AORI} at the University of Tokyo conducted a high-resolution \ac{MCS} reflection survey in March 2024, targeting the rupture zone of the 2024 Noto Peninsula earthquake. This study provides a detailed review of the survey and outlines the depth imaging workflow used in \ac{MCS} data processing, including velocity model building and seismic migration. Key challenges are also examined, particularly those associated with velocity model uncertainties, complex subsurface heterogeneity, and acquisition constraints. The resulting depth images contribute valuable structural information for understanding the rupture characteristics of this event. Finally, we discuss the implications of these findings for future surveys aimed at resolving outstanding questions related to crustal deformation, fault connectivity, and rupture propagation in this tectonically complex back-arc environment.

\section*{Data Acquisition Survey}
The MCS reflection survey was conducted aboard the R/V Hakuho-Maru from March 4 to 16, 2024, aiming to obtain a high-resolution image of the shallow crustal structure off the northeastern coast of the Noto Peninsula. Figure \ref{Area} presents a schematic map of the survey layout, which includes 14 2D seismic profiles, each approximately 46 km long. The survey parameters are summarized in Table \ref{tab:data-acquisition}.

\begin{table}[h!]
	\centering
	{\small 
		\begin{tabular}{|p{3cm}|p{3cm}|}
			\hline
			\multicolumn{2}{|c|}{\textbf{Data acquisition parameters}} \\ \hline
			\textbf{Parameter} & \textbf{Value} \\ \hline
			\small
			Receiver interval & 25 m \\ \hline
			Shot interval & 18.75 m \\ \hline
			Recording length & 5--6 seconds \\ \hline
			Seismic profile length & $~45$ km \\ \hline
			Sampling rate & 2 ms \\ \hline
			Number of Channels & 48 \\ \hline
			Maximum offset & 1284 m \\ \hline
			Source type & Two GI guns (355 x 2 = 710 cubic inch)  \\ \hline
			
		\end{tabular}
	} 
	\caption{Summary of data acquisition parameters of the multi-channel seismic (MCS) reflection survey conducted using the R/V Hakuho-Maru from March 4 to 16, 2024, in the northeast coast of the Noto Peninsula. }
	\label{tab:data-acquisition}
\end{table}

\FloatBarrier

\section*{Data Processing Workflow}

Seismic profiles were designed to cover the rupture area in the northeastern Noto Peninsula, with line orientations perpendicular to the rupture trajectory. The recorded data were processed using a two-stage workflow: 1) Pre-processing and 2) Advanced Depth Imaging. These steps were carried out using the commercial software RadexPro® and Paradigm®.

\subsection*{pre-processing}

In the pre-processing stage, navigation geometry was applied to the raw seismic shots, followed by binning with a bin size of 12.5 meters, equivalent to half the receiver spacing. This stage included deghosting \citep{berryhill1984wave} to enhance temporal resolution by mitigating ghost effects, as well as \ac{SRMA} \citep{verschuur1992adaptive} to suppress strong seafloor reflection multiples. Predictive deconvolution \citep{robinson1967predictive} was then employed to attenuate residual multiples, followed by velocity analysis, stacking, post-stack time migration, and time-to-depth conversion.

\subsection*{Advanced Depth Imaging}
A significant challenge in the depth imaging of the Noto dataset lies in building an optimal method for refining the interval velocity model for \ac{PSDM}. This difficulty arises due to the complex geological structures and, in particular, the absence of distinguishable, continuous reflectors, which limits the effectiveness of conventional layer-stripping techniques \citep{shih1996iterative}.

To address this challenge, we concentrated on extracting horizons identified by automatic picking methods. The process begins with the conversion of stacking velocities, obtained from the velocity analysis step, into a depth interval velocity section. We applied  a strong smoothing operator on the initial  stacking velocities to remove the adverse bumpy effect due to low fold coverage in the velocity analysis step.   The first step of Kirchhoff prestack depth migration (\ac{KPSDM}) is then performed using this interval velocity section. This method is widely used in the imaging of \ac{MCS} marine data \citep{holbrook1994deep,fliedner2003depth}.

Subsequently, the depth-migrated sections undergo additional processing to facilitate the automated extraction of dip and continuity attributes \citep{chopra2005seismic}. To suppress artifacts associated with the water column, a seafloor top mute is applied to the migrated sections before attribute estimation. An automatic gain control (\ac{AGC}) filter is also applied to enhance weak continuous reflectors, improving their visibility and amplifying their contribution to subsequent analysis. Structural attributes, such as dip and continuity, are then extracted from the processed migrated sections. These attributes are combined for the automatic picking of continuous horizons, identified through the creation of pencils \citep{bacon20073}.

A low thresholding parameter is applied to ensure maximal coverage in identifying continuous horizons, albeit at the expense of picking additional non-real horizons and increasing the risk of amplifying multiples. To address this issue, the generated pencils undergo quality control based on a list of Moveout Groups. Metrics such as Autopick Semblance, Residual Moveout, Parametric Semblance, and Stack are analyzed to distinguish primary reflections from multiples, as well as true horizons from isolated spikes or noise \citep{woodward1998automated, decker2022variational}. Based on this analysis, various Moveout Group masks are integrated to define the selected horizon segments, which are then used as input for the grid-based tomography step.

\begin{figure*}[h!]
	\centering
	\includegraphics[width=0.8\textwidth]{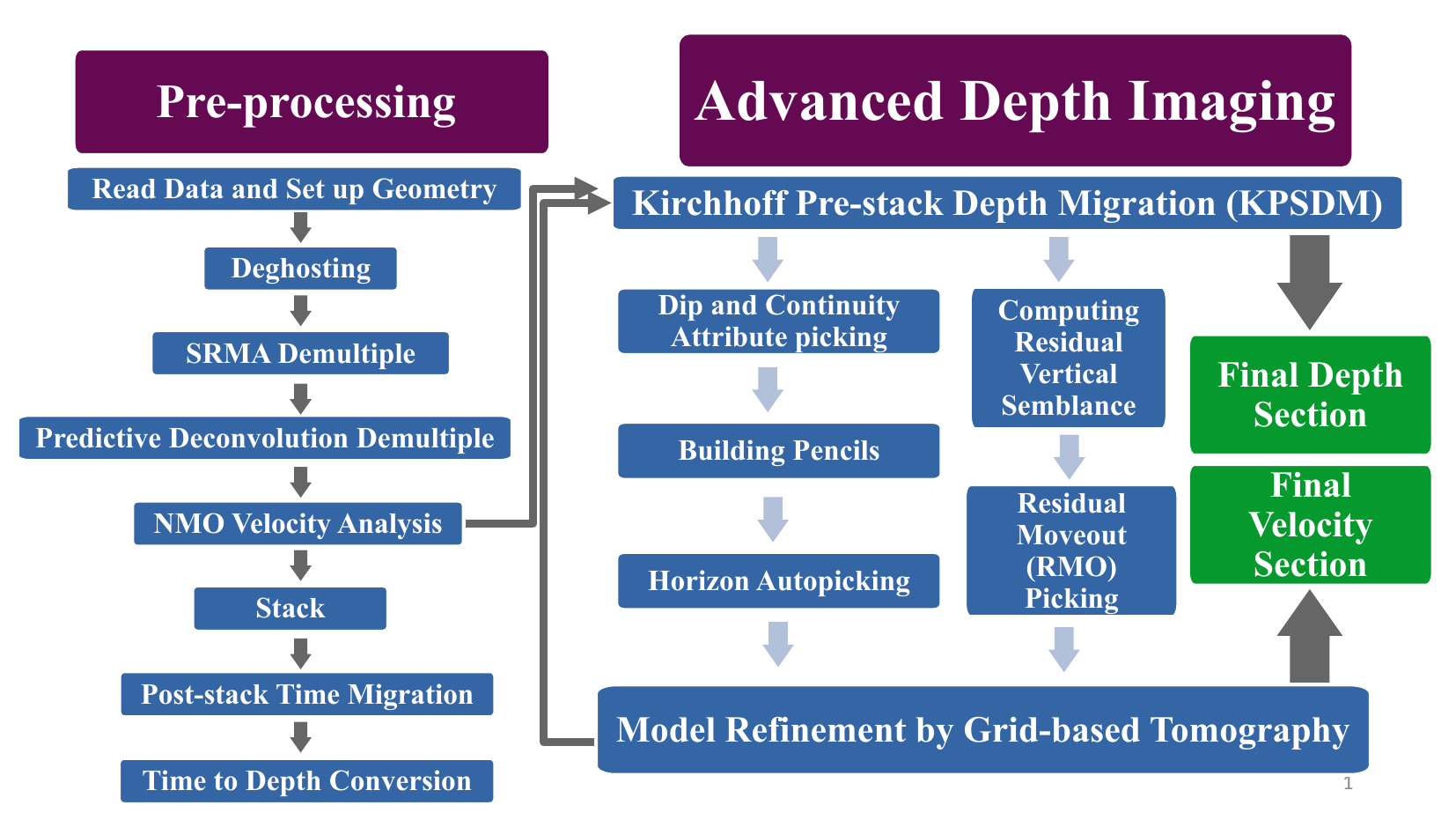}
	\caption{Two-step seismic processing workflow consisting of (1) Pre-processing and (2) Advanced Depth Imaging. The pre-processing stage includes navigation geometry definition, deghosting, surface-related multiple attenuation (SRMA), predictive deconvolution, velocity analysis, stacking, post-stack time migration, and time-to-depth conversion. The advanced depth imaging stage begins with an initial Kirchhoff pre-stack depth migration (KPSDM), followed by grid-based reflection tomography to refine the velocity model. Structural attributes, such as dip and continuity, are used for automated horizon picking through pencil structures, which undergo quality control using predefined Moveout Groups. An iterative tomography approach further updates the P-wave velocity model, enhancing seismic horizon alignment and minimizing residual moveout (RMO) in migrated gathers. In the final stage, manual RMO picking from residual semblances is incorporated into the tomography workflow to correct automated picking errors, ensuring additional quality control. This approach enables progressive velocity model refinement, leading to high-resolution depth-migrated sections and a final optimized velocity model. }
	\label{workflow}
\end{figure*}

The process continues with an iterative, grid-based tomography approach to further refine the P-wave velocity model, enhancing the alignment of seismic horizons and minimizing the \ac{RMO} in the calculated semblances of migrated seismic gathers \citep{luo2014imaging}. We conducted 4–5 rounds of grid-based tomography following the autopicking of horizons. In the later stages, the \ac{RMO} of migrated gathers is manually picked from residual semblances and integrated into the grid-based tomography. This manual step ensures that errors or ambiguities in the automated picks are addressed, providing an additional layer of quality control. This approach allows for gradual and controlled improvements in the velocity model, ultimately leading to high-resolution depth-migrated sections and the final velocity model [For more details, see \citep{hokstad2000multicomponent}]. A full review of the implemented workflow is presented in Diagram \ref{workflow}.

\FloatBarrier    

\FloatBarrier  

\vspace{-0cm}

\section*{Results and Discussion}

The rupture zone of the 2024 $M_w$ 7.6 earthquake, northeast of the Noto Peninsula, is expected to exhibit an intricate \ac{3D} subsurface structure due to the region’s tectonic regime, history of crustal deformation, complicated fault system (\ac{JSPJ} model; \cite{fujii2024slip}), and complex stress distribution. The results presented in this section were obtained using advanced depth imaging to optimally compile a \ac{2D} acoustic response of the complex \ac{3D} subsurface structure and investigate the impact of the rupture zone on intersecting \ac{2D} seismic profiles. The geological interpretations of these sections have been presented in \cite{park2026tsunamigenic}

\subsection*{Velocity Model Refinement}

\begin{figure}[h!]
	\centering
	\includegraphics[width=0.9\textwidth]{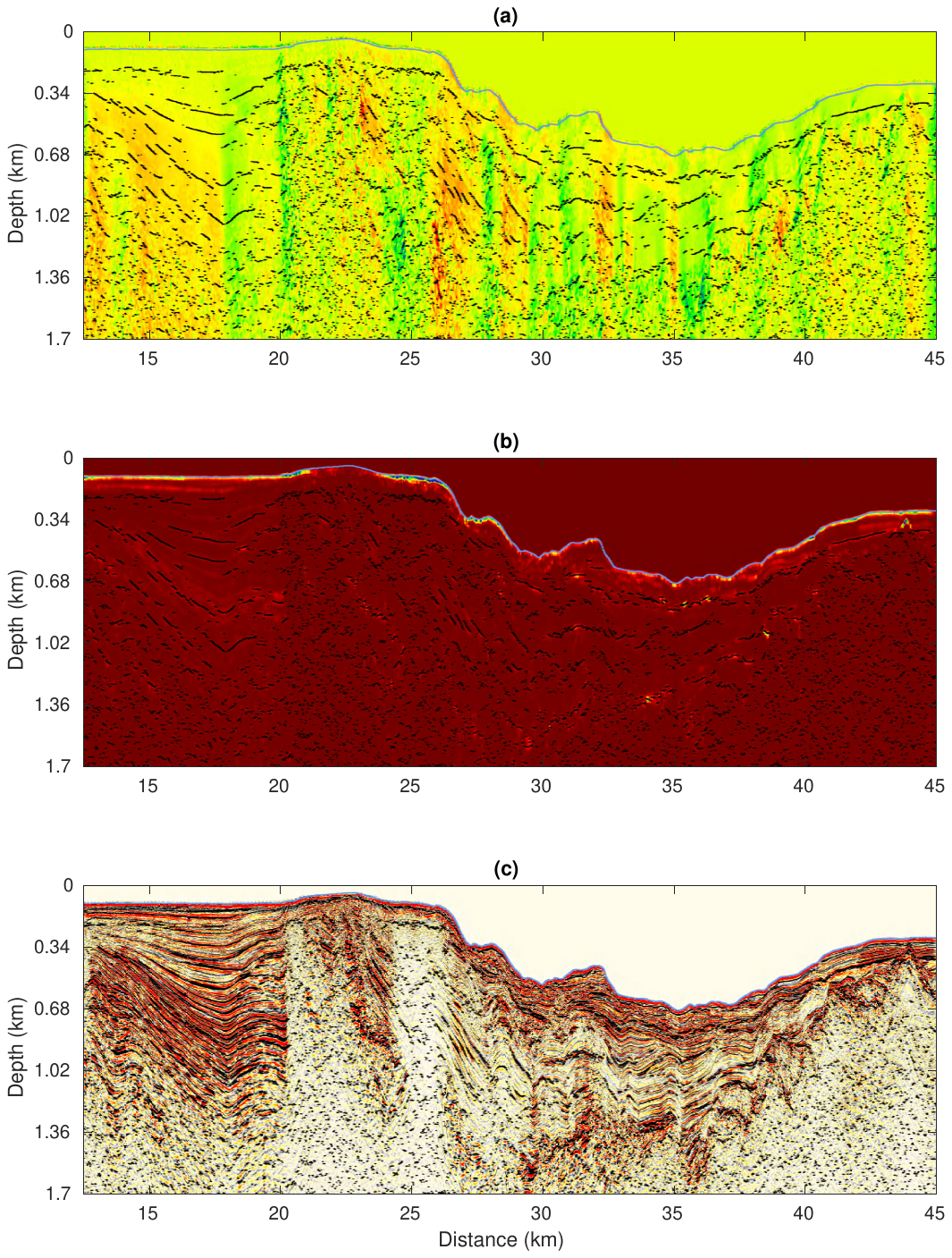}
	
	\caption{This figure illustrates automatic horizon picking using pencil structures. The (a) and (b) panels display the automatically extracted dip and continuity attributes, which serve as inputs for pencil picking. The black lines in all sections represent the picked pencils, generated by autopickers integrating dip and continuity attributes. The  panel (c) presents the input PSDM section for comparison and correlation assessment. As shown, many horizons have been accurately identified by the autopickers (see text for details).}
	\label{attr}
\end{figure}

We present the Advanced Depth Imaging component of our workflow (Fig.~\ref{workflow}), as applied to a representative 2D seismic line. Line K1 was selected for its distinctive characteristics, notably its relatively high signal-to-noise (S/N) ratio compared to other lines in the survey. Furthermore, the dataset for K1 is complete, with no missing shot records, thereby preserving the integrity of the imaging process.

A central step in this workflow is the application of horizon autopicking for velocity model refinement via grid-based tomography. A detailed illustration of this process is provided in Fig.\ref{attr}. Panels(a) and (b) display the dip and continuity attributes computed from the PSDM section of Line K1. For improved interpretability, a zoomed-in window is shown, covering a depth range of 0--1.7km and an inline range of 12--45km from the survey origin. All attribute values were normalized to the range [0,~1] to ensure consistent visual scaling without requiring color bars.

Black curves overlaid on all panels represent pencil structures generated from the dip and continuity attributes. These structures serve as the initial set of candidate horizons, which are subsequently refined using additional structural attributes, including Autopick Semblance, Residual Moveout, Parametric Semblance, and Stack. Panel~(c) displays the input PSDM section for reference, enabling visual correlation. A direct comparison between the autopicked pencil structures and the PSDM section confirms that major seismic horizons, although locally limited, have been accurately delineated. This result highlights the efficacy of structural attribute-guided horizon picking in supporting velocity model updates.

Dip and continuity attributes play a crucial role in reliable horizon autopicking by enhancing reflector identification and continuity. The dip attribute improves the detection of steeply dipping reflectors, which are often poorly imaged by PSDM due to Kirchhoff migration's limitations in resolving such features. This effect is particularly evident in the 25–30 km inline range, where reflectors appear weak in the PSDM section but are successfully highlighted through dip attribute integration, demonstrating its effectiveness in imaging steeply dipping geological structures. 

The continuity attribute, as shown in panel (b) of Fig. \ref{attr}, enhances the selection of high-amplitude, laterally continuous horizons. These horizons are naturally incorporated into the autopicked pencils, reinforcing the significance of continuity attributes in the horizon picking process.

Another critical factor in our processing workflow is the use of an extremely low threshold parameter during the initial pencil selection, specifically adapted to the dataset’s characteristics. Given the low fold coverage, short streamer length, and lack of distinct formation boundaries in the study area, a lower threshold enhances the detection of weak and discontinuous horizons. While this increases the risk of false horizon picks, as shown in panel (c) of Fig. \ref{attr}, it ensures a more comprehensive initial capture of geologically significant reflectors.To refine the selection, the generated pencils undergo quality control based on Moveout Groups. Additional attributes, including Autopick Semblance, Residual Moveout, Parametric Semblance, and Stack, are manually thresholded to distinguish primary reflections from multiples. This filtering process isolates a final set of horizons, which are then optimized using grid-based tomography.

Adapting to these settings, we perform 4--5 rounds of grid-based tomography following this strategy. To further refine the velocity model, we enforce the alignment of seismic horizons while minimizing \ac{RMO} in the calculated semblances of migrated seismic gathers. In the later stages, \ac{RMO} is manually picked from the residual semblance panels of the migrated gathers and incorporated into the grid-based tomography workflow.

\begin{figure}[h!]
	\centering
	\includegraphics[width=0.9\textwidth]{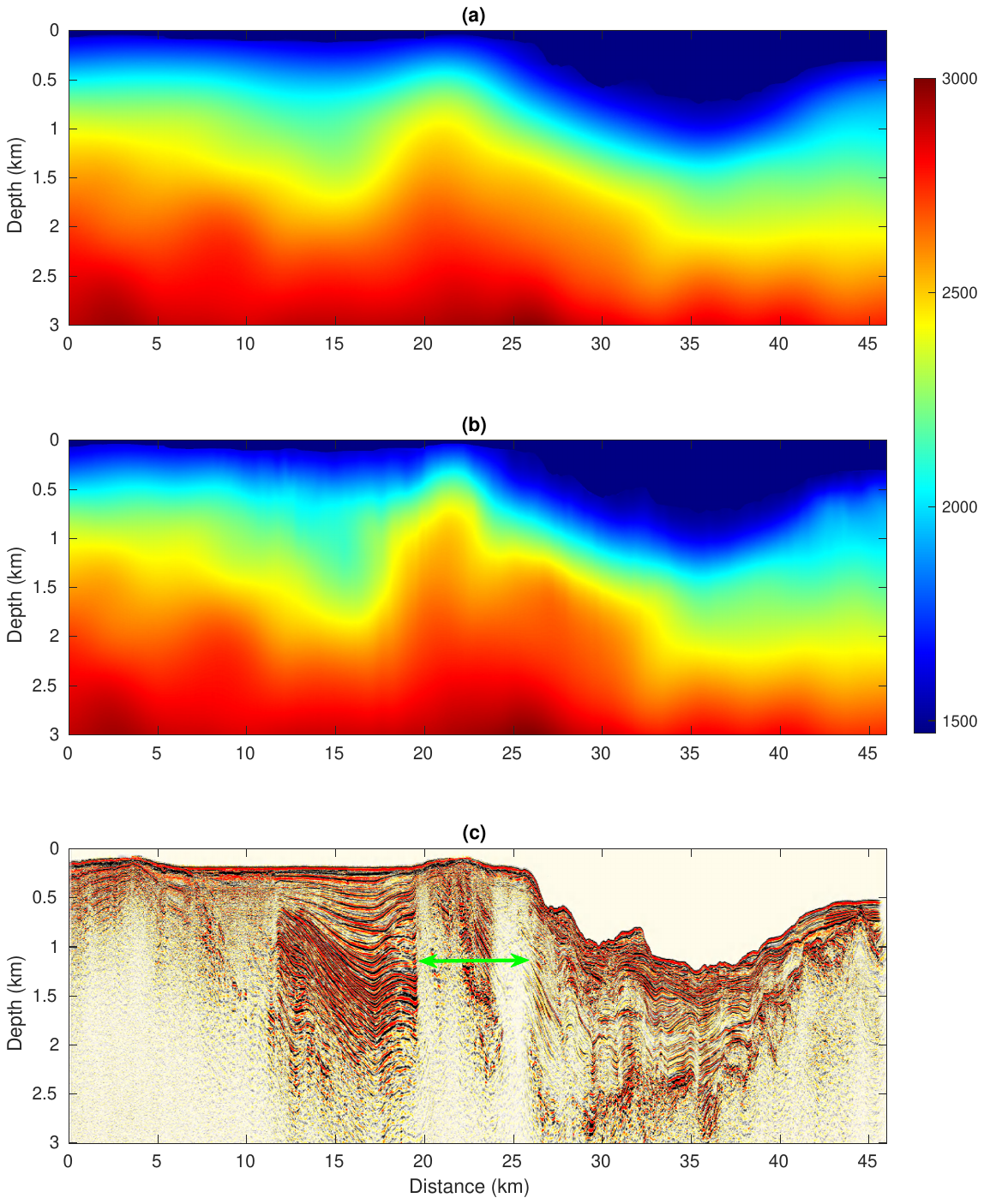}
	
	\caption{Comparison of the initial P-wave depth interval velocity model, derived from stacking velocities (panel (a)), and the final velocity model (panel (b)) for the line K1 obtained by the workflow in Fig. \ref{workflow}. The initial model has been updated using grid-based tomography, based on a combined integration of autopicked attributes and Migration Residual Moveout [see the text for more details]. A significant modification in velocity is observed at shallow depths, while the effect of tomography at deeper depths is minimal. The obtained PSDM section built by the  final depth interval velocity model by the advanced depth imaging process is presented in the panel (c). A high velocity zone anomaly within the 20–23 km inline range coinciding with a poorly imaged zone in the PSDM seismic section (highlighted by a green double arrow) is observed, likely caused by intense fracturing related to the 2024 Noto earthquake.}
	\label{vels}
\end{figure}

Fig. \ref{vels} presents the initial velocity model from the pre-processing step (panel (a)) and the refined velocity model obtained after multiple iterations of grid-based reflection tomography (panel (b)). Both velocity models are displayed with a vertical exaggeration of 440\% to enhance the resolution of vertical heterogeneities. Panel (c) of this figure shows the final PSDM section, generated using the final interval velocity model, revealing key structural features that were previously less resolved. Only the upper 3 km of the velocity models and PSDM section are displayed, as the remaining 2 km of the model lacks sharp reflectors. The refined velocity model and the depth-migrated seismic section provide improved resolution of the subsurface structure, allowing for a more detailed analysis. The following section presents an interpretation of key velocity anomalies and their geological significance.

\FloatBarrier

Considering the final velocity model (Fig. \ref{vels}, panel (b)), obtained from grid-based tomography, and the PSDM section (panel (c)) of Line K1, we can delineate structural anomalies along the profile. The refined velocity model provides improved structural resolution, allowing for a more detailed characterization of subsurface features compared to the initial velocity model, which primarily represents a low-resolution background velocity structure. Two pronounced \ac{HVZ} anomalies are identified within the inline ranges of 20–23 km and 25–30 km, extending from approximately 500 m depth downward. The first \ac{HVZ} anomaly exhibits an almost vertical geometry, suggesting a steeply dipping structural feature, whereas the second anomaly displays a distinct inclination, dipping towards deeper stratigraphic units along the survey line.
Additionally, a laterally continuous horizon with velocities exceeding 2500 m/s is observed along the entire seismic profile, which is attributed to the basement. The two \ac{HVZ} anomalies may represent unconformities resulting from basement uplift within the 20–23 km and 25–30 km segments. These structural features could be indicative of tectonic reactivation processes, leading to localized uplift and deformation of the overlying sedimentary sequences.

These \ac{HVZ} correspond to poorly imaged regions in the PSDM section, suggesting a massive, homogeneous lithology with no sharp geological contrasts. Moreover, the HVZ anomaly within the 20–23 km inline range is particularly significant, as it coincides with a poorly imaged zone in the PSDM seismic section (highlighted by a green double arrow), likely caused by intense fracturing related to the 2024 Noto earthquake. This feature is clearly resolved in the K1 profile and appears consistently across adjacent lines from Line 9 to K1 (see subsections \ref{3-10} and \ref{11-k1}), indicating a laterally continuous tectonic structure. Its alignment with the estimated rupture area, as inferred from independent seismic data, supports its interpretation as a zone of localized deformation. The \ac{3D} visualization of the 2D seismic sections in subsection \ref{3Dsex} further underscores this spatial correlation, highlighting the anomaly's continuity within the rupture zone.

In contrast, a sharp \ac{LVZ} has developed between distance 8-20km inline within the 500–2000 m depth range. Counterintuitively, despite its low-velocity nature, the PSDM section in this interval exhibits sharp, continuous reflectors with a complex, intensely folded structure. A distinct boundary is observed in the PSDM section between the \ac{HVZ} and \ac{LVZ}, suggesting a significant structural or lithological transition. A second \ac{LVZ} is observed within the inljne distance 12-19 km at similar depths, likely emerging as a result of iterative velocity updates. These velocity variations underscore the effectiveness of grid-based tomography in refining the velocity structure and improving subsurface characterization.

\subsection*{Compiled Depth Sections}
\subsubsection*{Lines 3-10}  \label{3-10}
\begin{figure}[H] 
	\centering
	\begin{subfigure}[t]{0.48\textwidth}
		\centering
		\includegraphics[width=\textwidth]{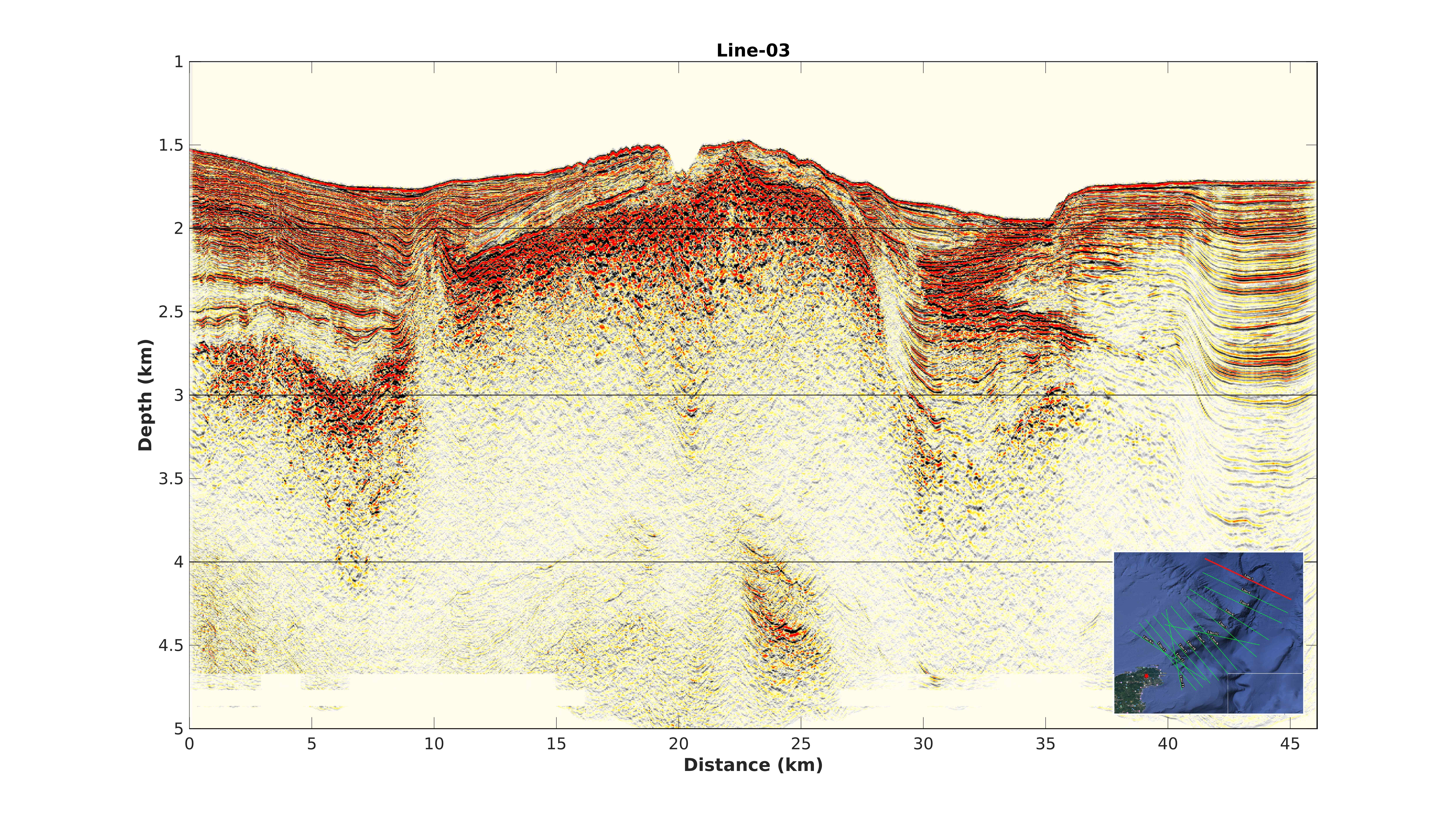}
		\caption{Line 03}
		\label{fig:03}
	\end{subfigure}
	\hfill
	\begin{subfigure}[t]{0.48\textwidth}
		\centering
		\includegraphics[width=\textwidth]{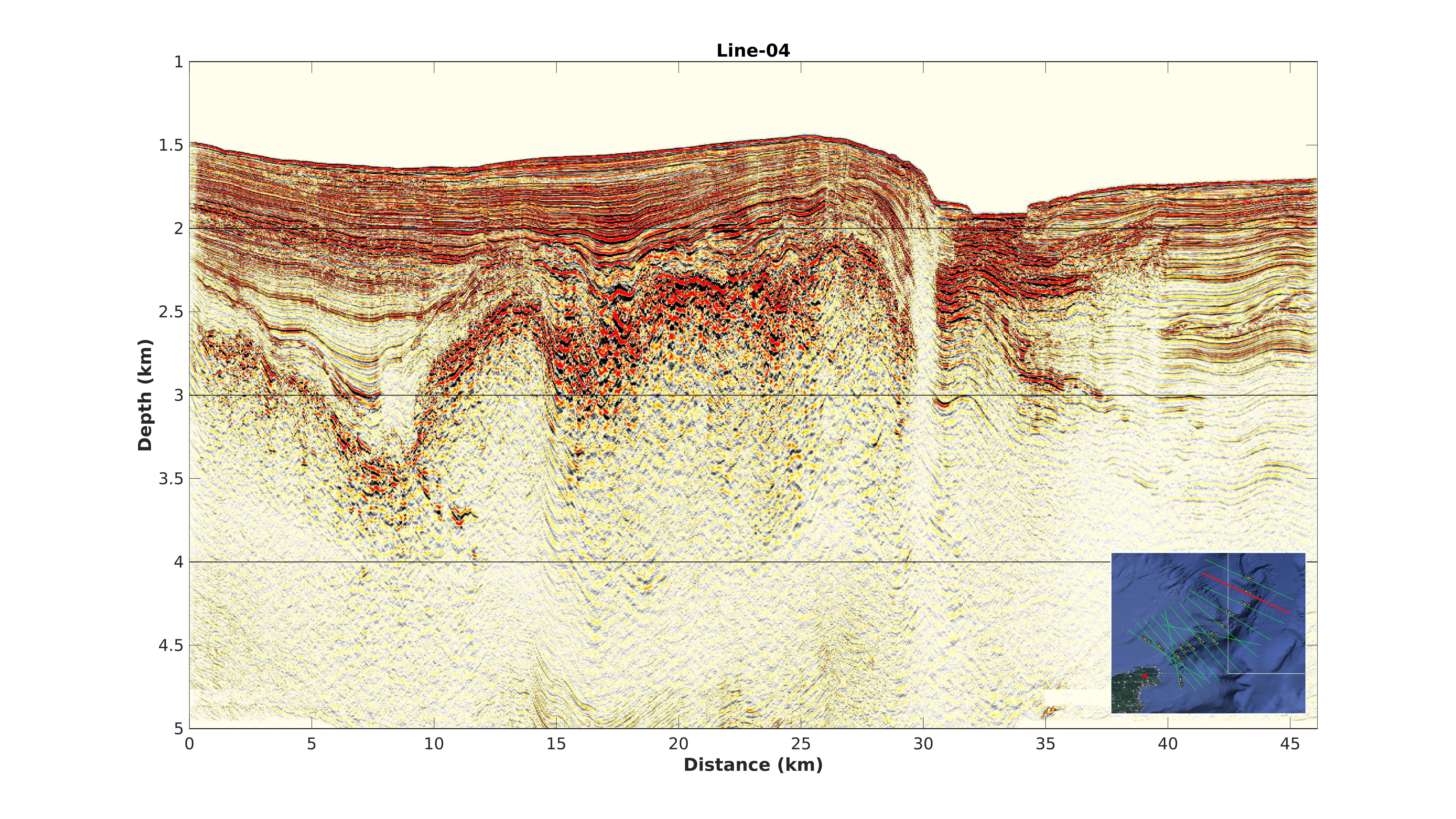}
		\caption{Line 04}
		\label{fig:04}
	\end{subfigure}
	
	\vspace{0.5em} 
	
	\begin{subfigure}[t]{0.48\textwidth}
		\centering
		\includegraphics[width=\textwidth]{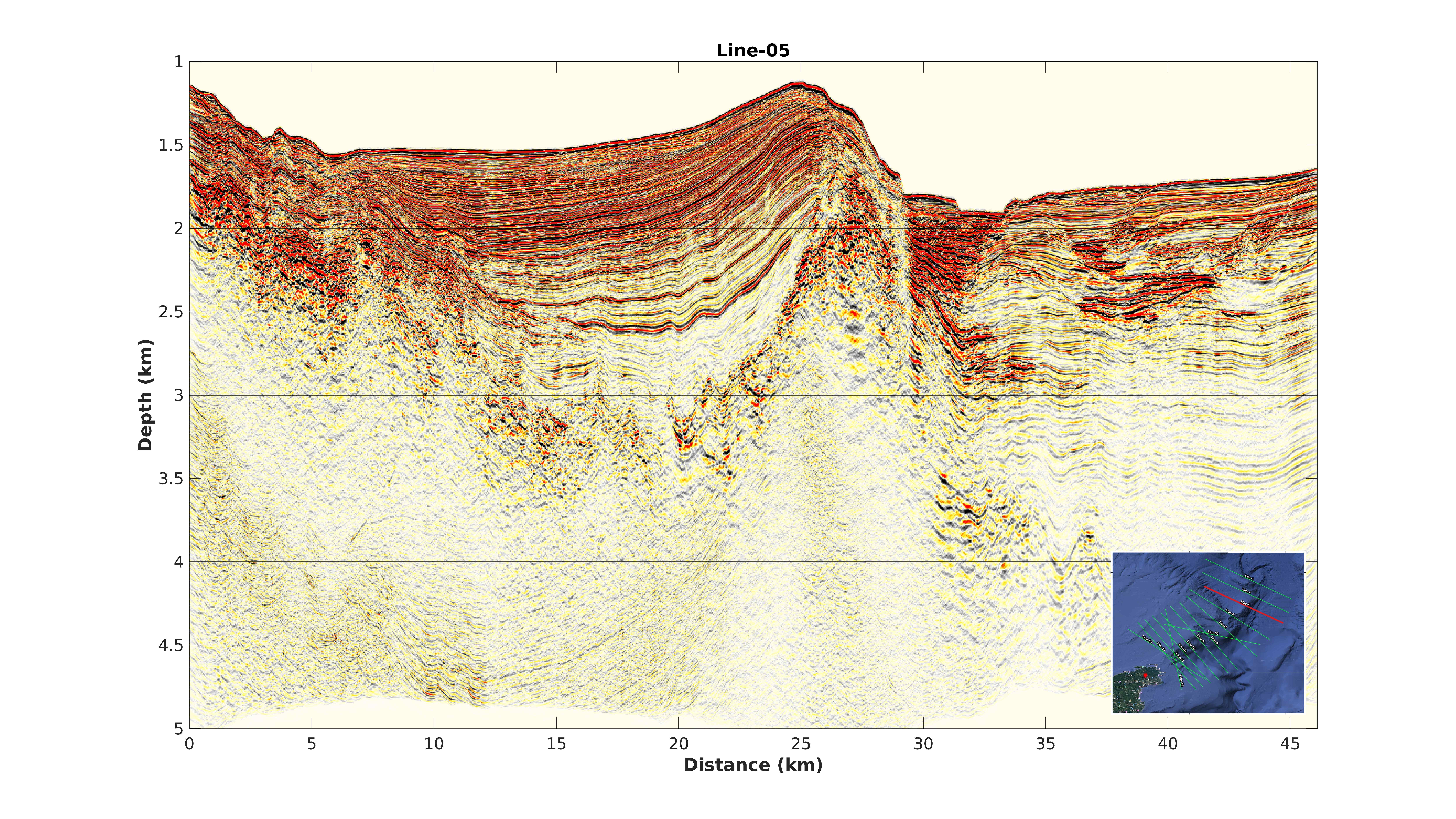}
		\caption{Line 05}
		\label{fig:05}
	\end{subfigure}
	\hfill
	\begin{subfigure}[t]{0.48\textwidth}
		\centering
		\includegraphics[width=\textwidth]{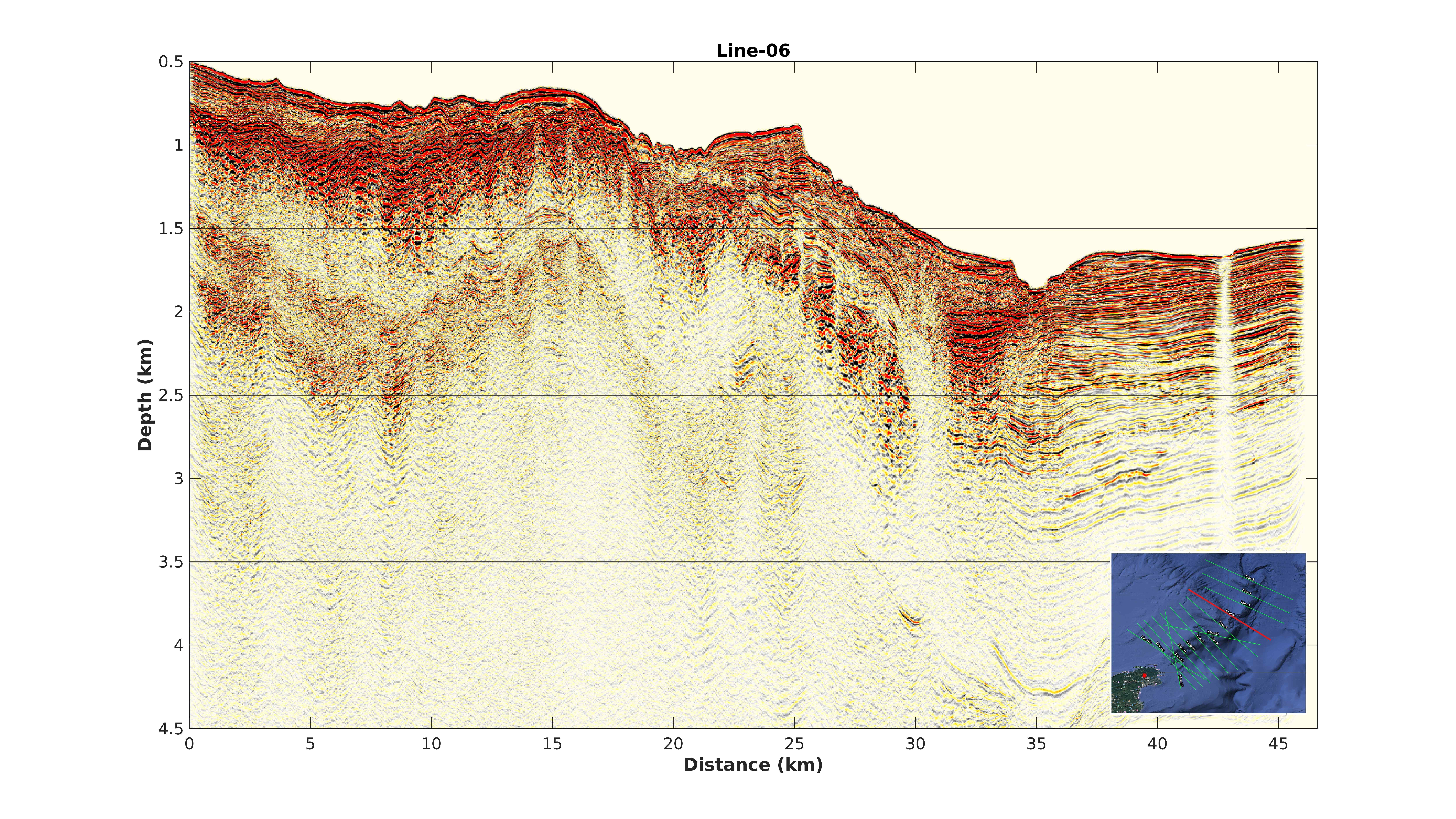}
		\caption{Line 06}
		\label{fig:06}
	\end{subfigure}
	
	\vspace{0.5em} 
	
	\begin{subfigure}[t]{0.48\textwidth}
		\centering
		\includegraphics[width=\textwidth]{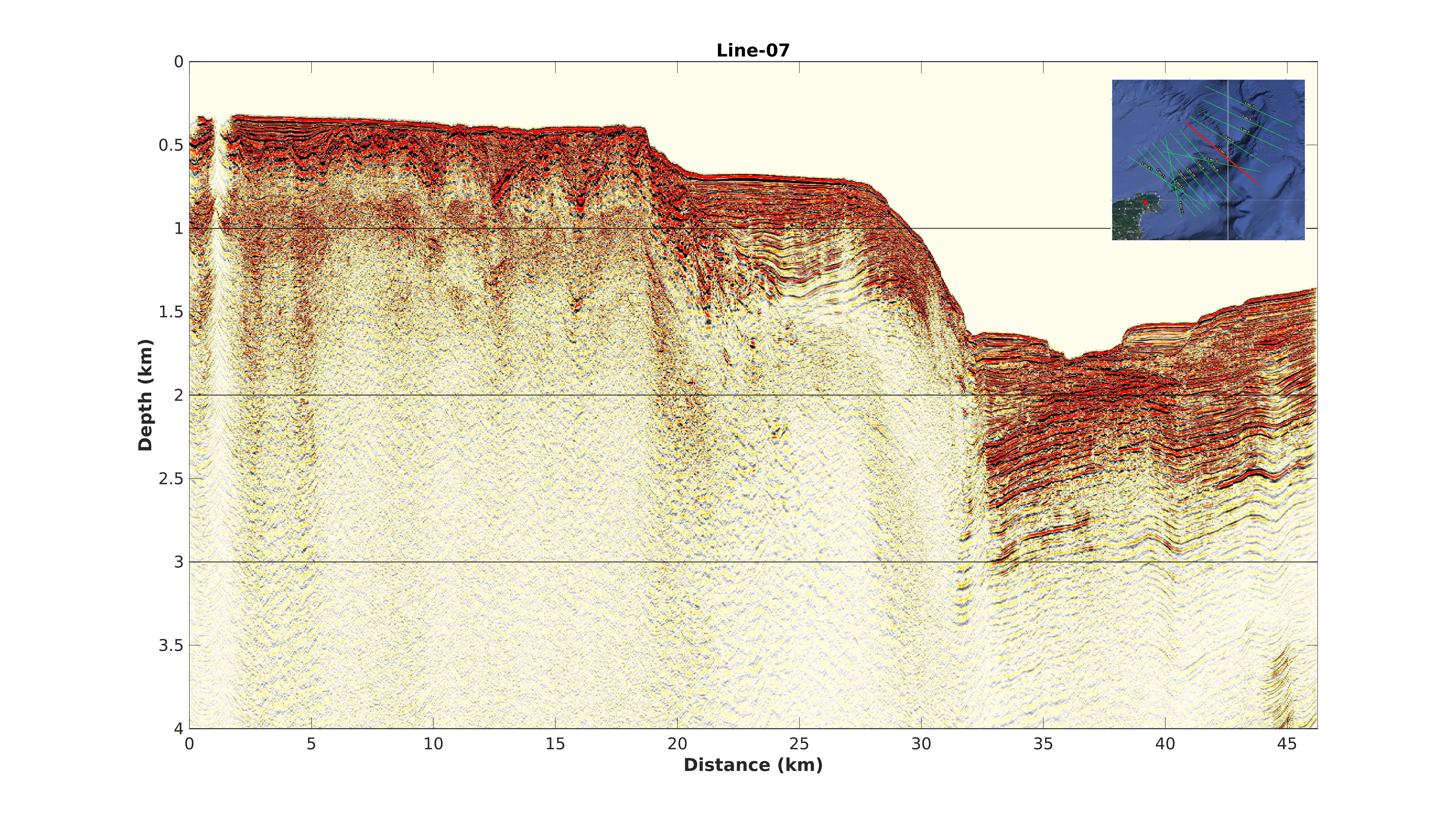}
		\caption{Line 07}
		\label{fig:07}
	\end{subfigure}
	\hfill
	\begin{subfigure}[t]{0.48\textwidth}
		\centering
		\includegraphics[width=\textwidth]{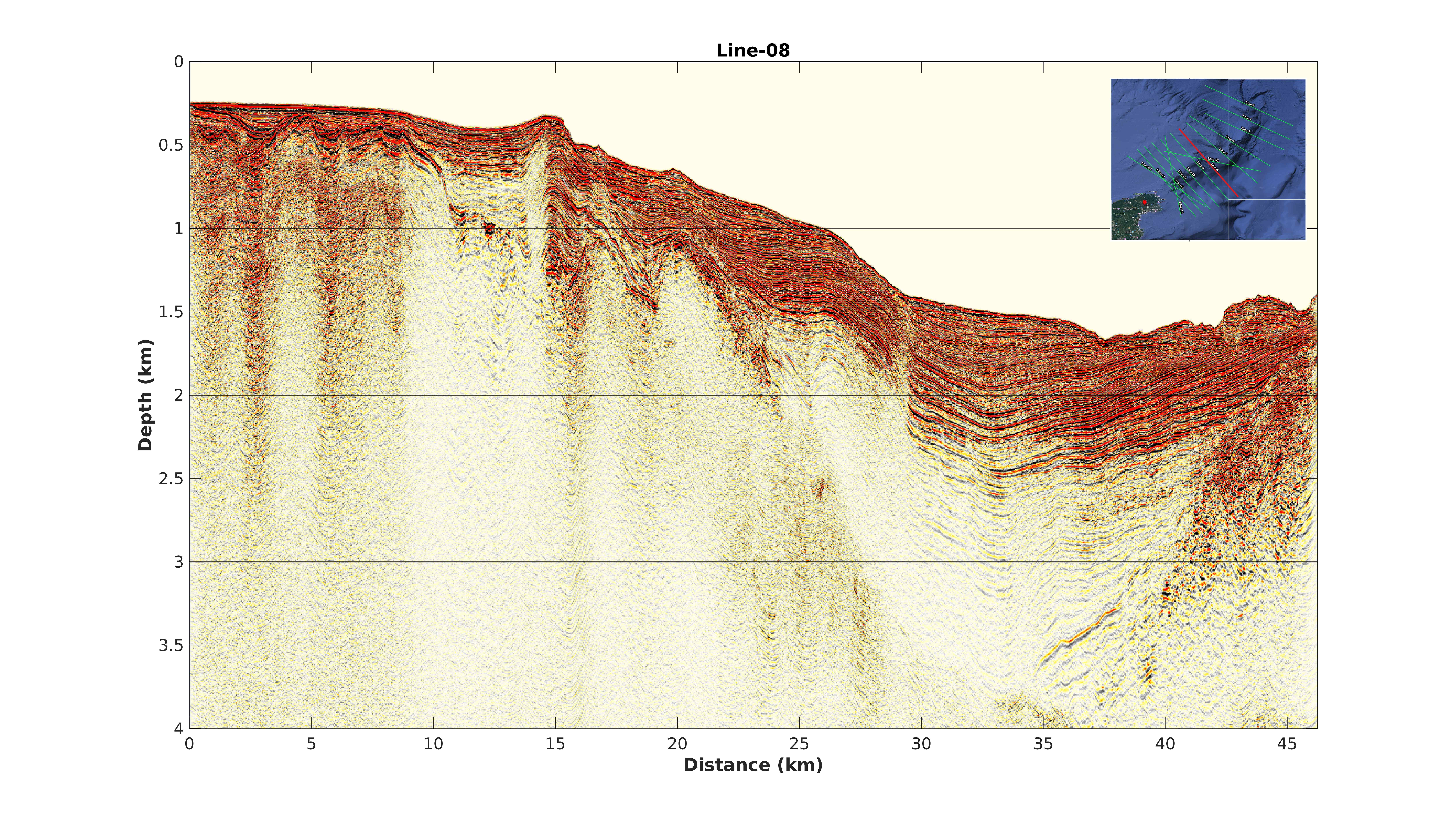}
		\caption{Line 08}
		\label{fig:08}
	\end{subfigure}
	
	\vspace{0.5em} 
	
	\begin{subfigure}[t]{0.48\textwidth}
		\centering
		\includegraphics[width=\textwidth]{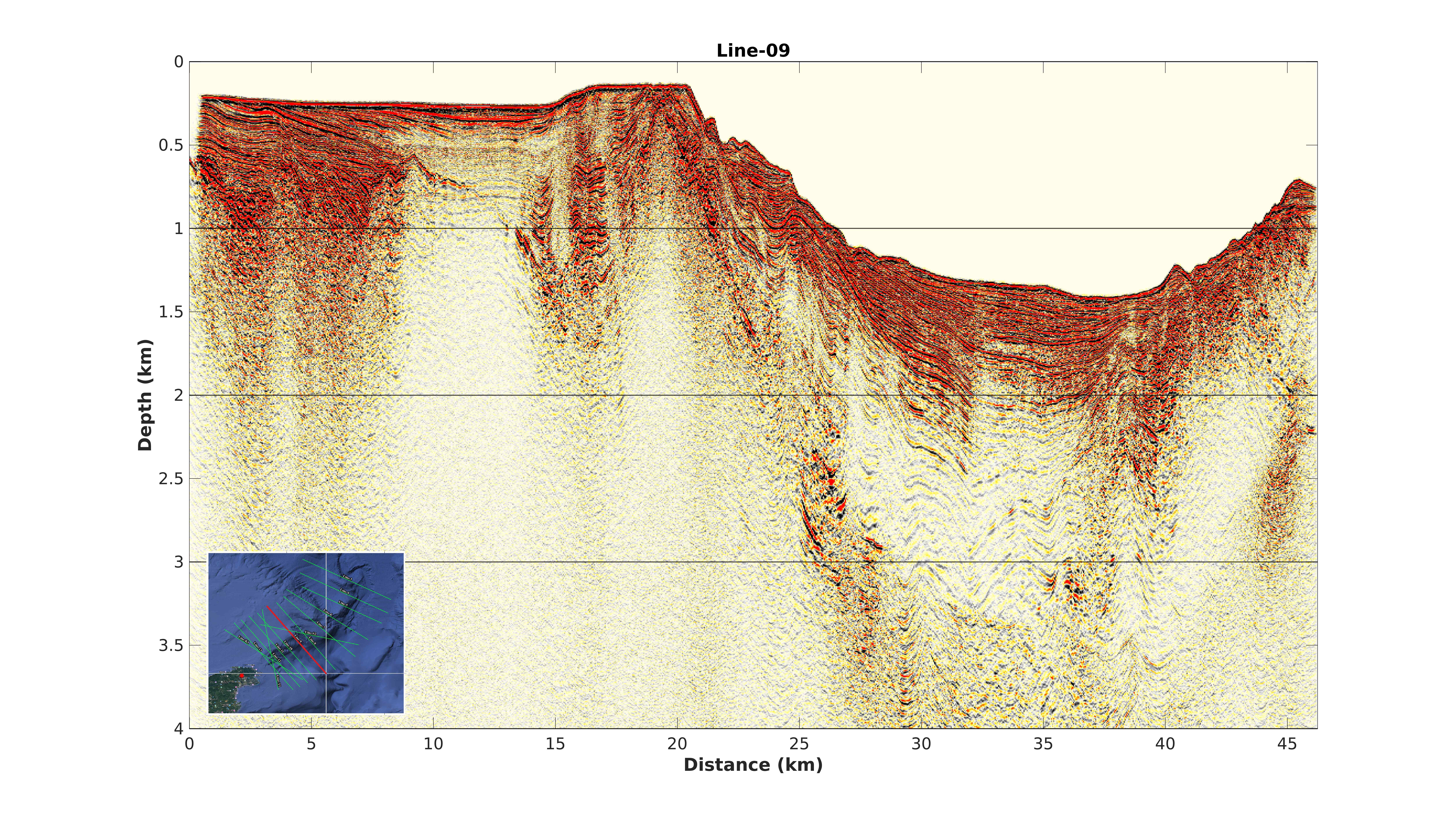}
		\caption{Line 09 }
		\label{fig:07_repeated}
	\end{subfigure}
	\hfill
	\begin{subfigure}[t]{0.48\textwidth}
		\centering
		\includegraphics[width=\textwidth]{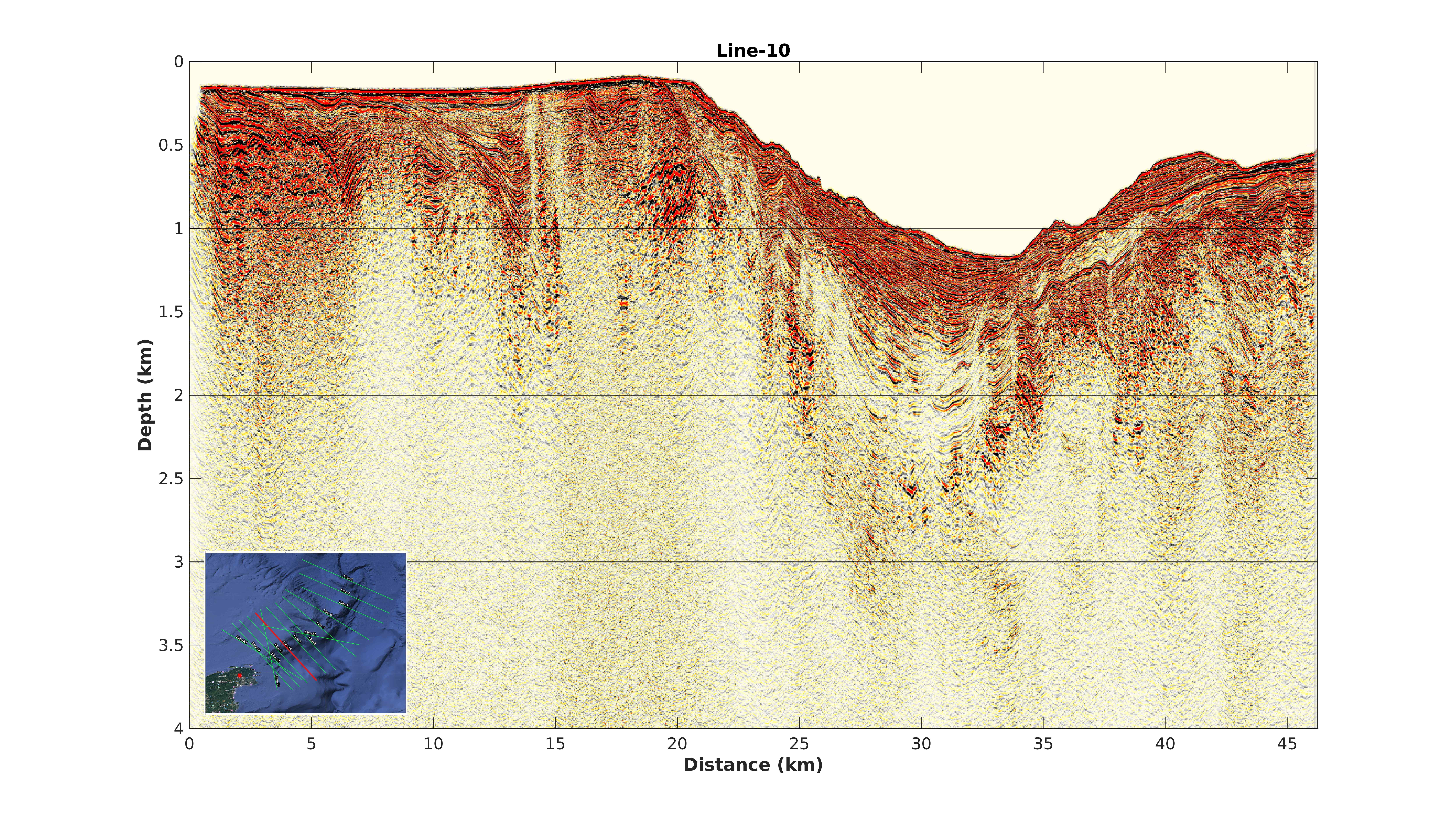}
		\caption{Line 10 }
		\label{fig:08_repeated}
	\end{subfigure}
	
	\captionsetup{justification=centering} 
	\caption{Pre-stack depth-migrated (PSDM) sections for Lines 03, 04, 05, 06, 07, 09, and 10. The geographical positions of these lines are highlighted in red among other lines plotted in green. The  \ac{2D} seismic section is displayed with a vertical exaggeration of 5x to enhance the visualization of subsurface features. High-resolution images of each section are provided in the supplementary material of the report. }
	\label{sections_3-10}
\end{figure}

In this section, we present the final compiled \ac{2D} PSDM sections, obtained following the processing workflow outlined in Fig. \ref{workflow}. Fig. \ref{sections_3-10} displays the \ac{KPSDM} sections for Lines 03, 04, 05, 06, 07, 09, and 10, arranged sequentially from the northeastern part of the rupture zone to the approximate middle-southern segment. A key characteristic of these lines is their orientation, as they are positioned nearly perpendicular to the rupture zone, providing optimal imaging of fault structures and deformation patterns. To aid spatial interpretation, each subfigure includes an inset map illustrating the geographical position of the corresponding seismic line. The analyzed profile is highlighted in red, while other seismic lines in the region are shown in green for reference. This visualization enhances the ability to correlate seismic sections with regional structural trends.

\subsubsection*{Lines 11, 12, 13, S2, T1, and K1} \label{11-k1}

Figure \ref{sections_11-K1} presents the \ac{KPSDM} sections for Lines 11, 12, 13, S2, T1, and K1, arranged from the middle-southern part of the survey area to the southwestern segment. Unlike the other profiles, the latter three lines (S2, T1, and K1) are not oriented perpendicular to the rupture zone. Instead, they are positioned to intersect other seismic lines, allowing for section correlation and structural continuity assessment. 	To enhance the visualization of subsurface features, a uniform vertical exaggeration factor of 4.4× was applied to all \ac{2D} seismic sections. High-resolution versions of each section are provided in the supplementary material, offering greater detail for in-depth structural analysis.

\subsection*{3D section plots of the depth sections} \label{3Dsex}

Advances in seismic interpretation emphasize the importance of \ac{3D} visualization for improving geological understanding in complex tectonic settings. Numerous studies have demonstrated that integrating 2D seismic profiles within a \ac{3D} spatial context significantly enhances the accuracy of structural and stratigraphic interpretation, especially in areas with intricate fault systems or folded strata \citep{Brown2011, Yilmaz2001, Wu2018}. Unlike isolated 2D sections, \ac{3D} visualization enables interpreters to view geometries and relationships in their true spatial orientation, reducing ambiguities caused by projection effects, line spacing, or oblique structural features. This approach has proven particularly valuable in delineating fault connectivity, resolving horizon terminations, and identifying features such as en echelon folds, branching faults, and stratigraphic pinch-outs, which are difficult to fully capture in single-section analysis. Additionally,  \ac{3D} interpretation supports better correlation across intersecting lines, leading to improved geological models and more robust hazard assessments in tectonically active regions \citep{Bond2007, Rowland2015}.

\begin{landscape}
	\begin{figure}[p]
		\centering
		\includegraphics[width=1.4\textheight]{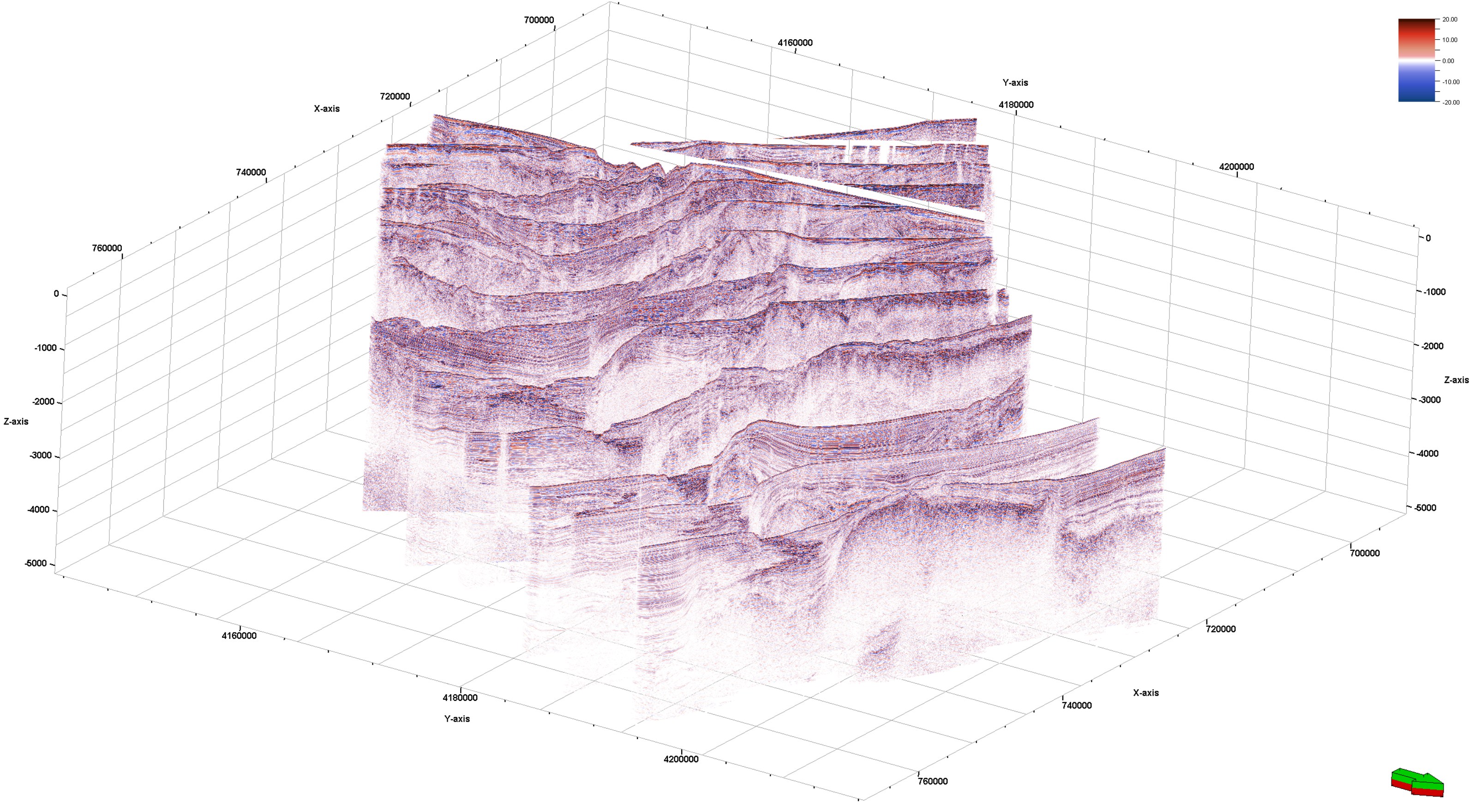}
		\caption{3D visualization of the integrated \ac{2D} seismic depth sections acquired in this study. Each profile is positioned according to its true spatial location, based on the conversion of Common Depth Point (\ac{CDP}) geographic coordinates to Universal Transverse Mercator (UTM) coordinates using the WGS 1984 datum in Zone 53N (135°E). The orientation arrow in the top-right corner indicates geographic north. This integrated \ac{3D} view enables structural features such as fault planes and stratigraphic horizons to be traced across intersecting sections, providing improved insight into the regional subsurface architecture.}
		\label{3Dmain}
	\end{figure}
\end{landscape}

Based on this rationale, the \ac{2D} depth sections generated in this study were integrated into a cohesive \ac{3D} spatial volume to enhance structural visualization and support detailed interpretation of the seismic dataset. Figure~\ref{3Dmain} presents the assembled seismic sections rendered within a \ac{3D} volume, providing a comprehensive overview of subsurface structural configurations across the survey area. The green-red directional arrow at the top left of the figure indicates the orientation of geographic north for spatial reference. To accurately geolocate the profiles in \ac{3D} space, the geographic coordinates of the \acp{CDP} were converted to Universal Transverse Mercator (UTM) coordinates using the WGS 1984 datum and UTM Zone 53N (central meridian at 135°E). For visualization clarity, the water column portion of each profile was removed, allowing clearer focus on sub-seafloor structures and facilitating more effective spatial interpretation.

\begin{figure}[H]
	\centering
	\begin{subfigure}[t]{0.75\textwidth}
		\centering
		\includegraphics[width=\textwidth]{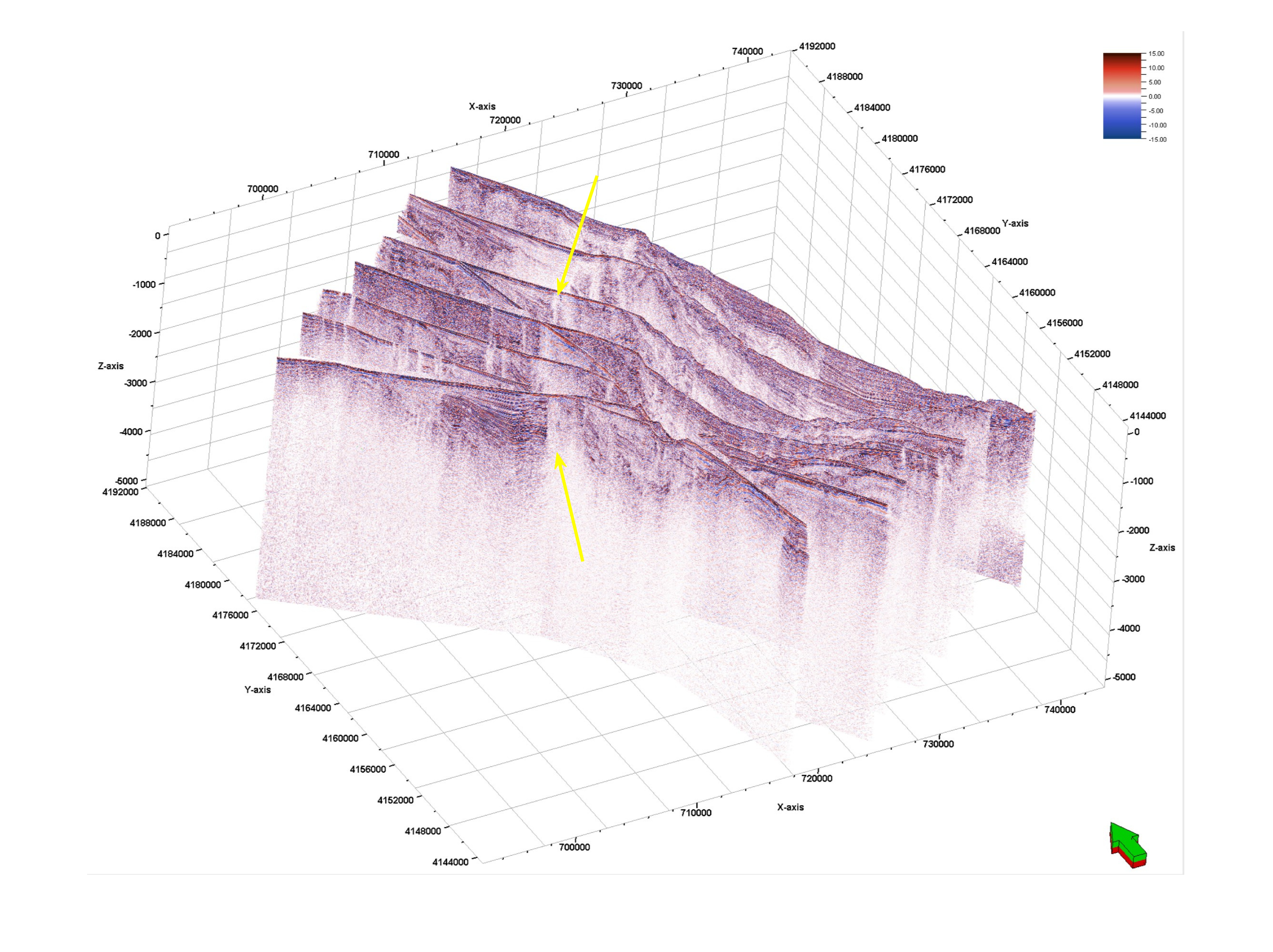}
		\caption{Lines 8, 9, 10, 11, 12, and K1.}
		\label{fig:03}
	\end{subfigure}	
	\vspace{1em}
	\begin{subfigure}[t]{0.71\textwidth}
		\centering
		\includegraphics[width=\textwidth]{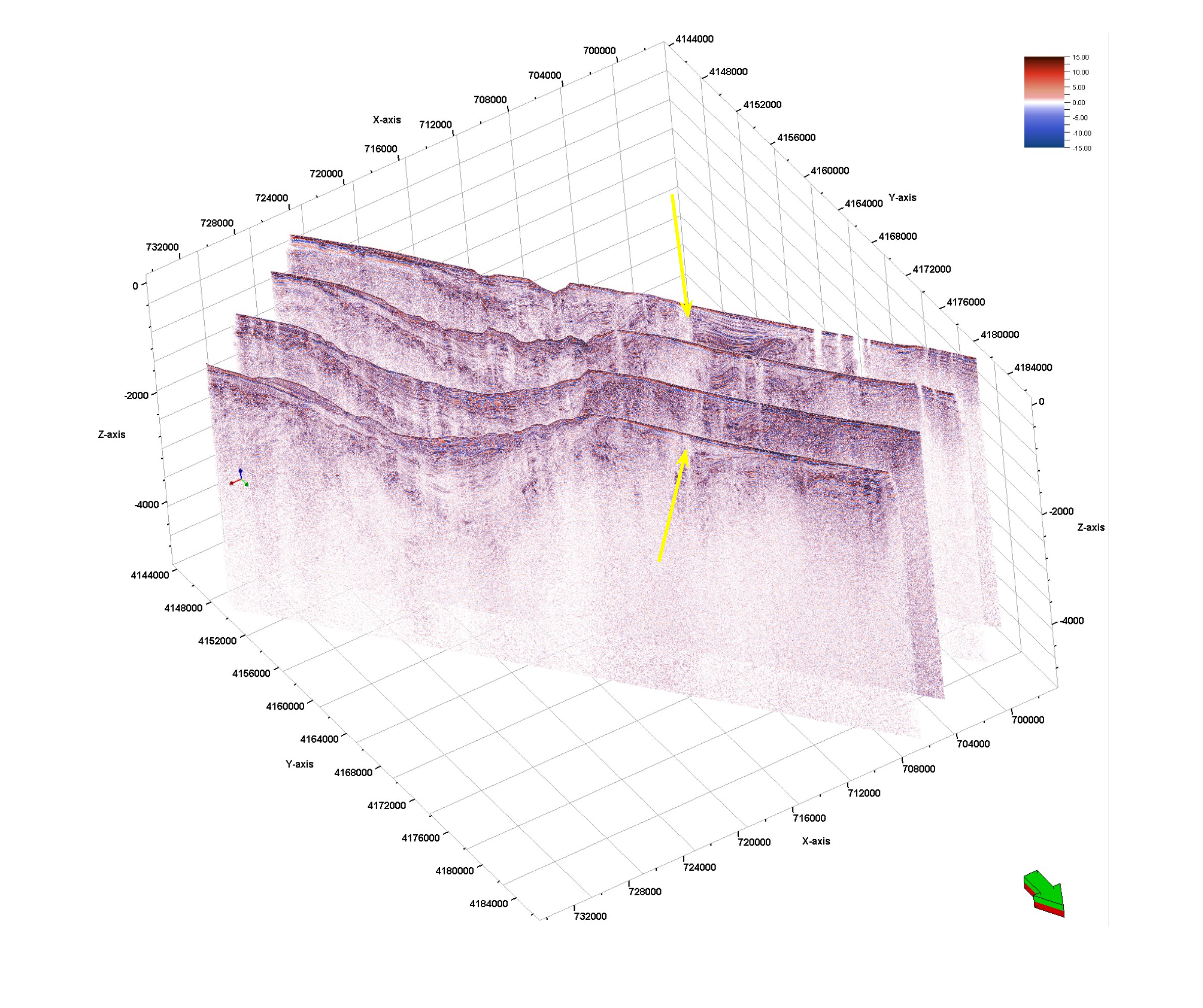}
		\caption{Lines 10, 11, and 12, from an opposing view.}
		\label{fig:05}
	\end{subfigure}
	\captionsetup{justification=centering}
	\caption{Focused \ac{3D} views of the rupture zone imaged in this study. Yellow arrows highlight rupture-related features traceable across intersecting seismic profiles. See also Figs.~\ref{sections_3-10} and \ref{sections_11-K1} for corresponding 2D sections.}
	\label{rupturee}
\end{figure}

Focusing on a specific region within the \ac{3D} seismic volume, we generated two data subsets that emphasize the portion of the imaged structure associated with the rupture zone. Both subfigures are spatially adjusted to provide optimal viewing orientation, clearly illustrating the interpreted rupture features within the \ac{3D} context, as shown in Fig.~\ref{rupturee}. Panel (a) delineates the rupture zone imaged along lines 8, 9, 10, 11, 12, and K1, while panel (b) presents the same rupture features from an opposing viewing angle, focusing on lines 10, 11, and 12. Two yellow arrows indicate the locations of the imaged rupture zone on the corresponding seismic sections. The rupture zone is also observable in the \ac{2D} depth sections presented in Figs.\ref{sections_3-10} and \ref{sections_11-K1}, specifically in panels (f), (g), and (h) of the former, and panels (a) through (f) of the latter. However, the rupture geometry and spatial extension is more clearly traceable in the \ac{3D} visualization shown in Fig.\ref{rupturee}, where spatial relationships and continuity are more readily apparent.

\clearpage  

\subsection*{limitations of data acquisition}

The resolution and interpretability of the seismic depth sections presented in this study are constrained by several inherent limitations in the data acquisition geometry. The primary factors include the relatively coarse receiver spacing, limited number of channels (48) in the recording streamer, and the restricted maximum offset associated with the short streamer length. These acquisition constraints directly impact the effectiveness of velocity model building and refinement, particularly in the context of grid-based tomographic inversion. Since the accuracy of tomographic updates depends on wide-angle coverage and offset diversity, the limited offset range reduces angular illumination, weakens velocity sensitivity at depth, and restricts the ability to constrain steeply dipping features.

The station spacing of 25 meters further limits lateral resolution by undersampling the recorded wavefield. In seismic imaging, finer receiver spacing contributes to improved signal coherence and allows for better reconstruction of complex subsurface geometries. Additionally, a higher trace density increases the signal-to-noise ratio (S/N) and improves the continuity and sharpness of imaged reflectors. In marine seismic acquisition, the fold number ($f_n$), which reflects the redundancy of ray paths contributing to each common midpoint (CMP), is given by
\[
f_n = \frac{S_l}{R_i},
\]
where $R_i$ is the station interval and $S_l$ denotes the offset range over which a given reflection is sampled \citep{yilmaz2001seismic}. In this survey, the combination of a 48-channel streamer and 25-meter station spacing resulted in a maximum fold of only 25, which is considered low for high-resolution prestack depth migration. Low fold coverage adversely affects the S/N ratio and undermines the stability of imaging, particularly in deeper sections where coverage is further reduced due to geometric spreading and attenuation.

While the applied imaging workflow enabled the recovery of major structural features and fault geometries, limitations in fold coverage and angular illumination hindered the resolution of fine-scale stratigraphic terminations and secondary fault branches. These limitations are particularly evident in zones of complex fault kinematics and near-surface heterogeneity, where imaging artifacts and loss of reflector continuity become more pronounced. Furthermore, \ac{2D} seismic profiling inherently assumes structural continuity within the inline plane, and therefore cannot adequately capture three-dimensional features such as out-of-plane fault segments, en echelon structures, or sideswipe energy. To overcome these limitations and improve characterization of rupture-related deformation, a dedicated 3D \ac{MCS} survey is necessary. High-density 3D acquisition would enable more complete spatial sampling of the wavefield, support improved velocity model building, and facilitate robust imaging of complex fault geometries within the shallow crust.

\section*{Conclusion and Future Work}

This study provides the first high-resolution seismic images of the shallow rupture zone associated with the 2024 $M_w$ 7.6 Noto Peninsula earthquake. Using a recently acquired MCS dataset, we applied an advanced depth imaging workflow to generate PSDM sections across 14 2D seismic lines intersecting the rupture area. The imaging procedure incorporated grid-based tomographic velocity model building, supported by automated picking of continuous reflectors guided by dip and coherency attributes, resulting in improved structural continuity and enhanced interpretability of the depth-migrated images.

The resulting seismic volume offers detailed 2D and 3D visualizations of the shallow crustal architecture within this tectonically inverted back-arc region. On the basis of these high-resolution images, we mapped the rupture zone across multiple intersecting profiles, allowing for spatial delineation of fault-related deformation and near-surface structural expressions of the causative fault. These observations provide new constraints on the geometry of the seismogenic fault system and its possible connection to the tsunami-generating segment inferred in previous studies. 

Despite the successful mapping of the rupture zone, the detailed structure of the fault is not adequately resolved due to limitations in the recorded dataset, primarily related to the large station spacing and the short streamer length used during acquisition. These constraints limited the spatial resolution and continuity of near-surface reflections, reducing the effectiveness of structural interpretation in key areas. Furthermore, \ac{2D} seismic profiling inherently assumes structural continuity within the inline plane and is therefore limited in its ability to resolve three-dimensional geological features. As a result, out-of-plane fault segments, en echelon fault arrangements, and sideswipe energy are often misrepresented or entirely missed in 2D imaging. These limitations significantly affect the interpretation of complex fault systems and shallow rupture processes, particularly in tectonically inverted settings like the Noto Peninsula.

Beyond the scope of the 2024 Noto event, this dataset constitutes a valuable geophysical benchmark for future investigations into active faulting, shallow rupture propagation, and tsunami generation processes along the eastern margin of the Sea of Japan. Given the limited availability of high-resolution seismic imaging in this region, our study fills a critical observational gap and contributes foundational data for advancing seismotectonic models in back-arc settings.

Although the present work focuses on imaging methodology with limited structural interpretation, the seismic sections reveal prominent fault geometries and stratigraphic terminations that correlate spatially with the inferred rupture trace. These features offer important insights into the mechanical segmentation and potential rupture pathways of offshore fault systems in compressional back-arc environments.

Future research will aim to integrate the MCS data with complementary geophysical and geological datasets, including high-resolution bathymetry, ocean floor video imagery, regional geologic maps, and geodetic measurements. Such integrated analyses will enable more comprehensive interpretations of fault zone architecture, fluid-driven deformation, and rupture dynamics. Further velocity model refinement, including full-waveform inversion and elastic imaging techniques, may improve resolution in the shallow crust, particularly in zones of low-velocity anomalies or complex structure. These efforts are expected to advance our understanding of shallow fault mechanics and tsunami generation processes, ultimately contributing to improved seismic hazard assessment along the Japan Sea margin.

\section*{Data availability}	
High-resolution depth-migrated seismic sections corresponding to all survey lines are available at \href{https://github.com/sigproseismology}{https://github.com/sigproseismology}. Additional datasets and processed seismic volumes used in this study are available from the corresponding author upon reasonable request.

\begin{acronym}[STA/LTA]
	\acro{STA/LTA}{short-term average over long-term average}
	
	\acro{DS}{deeply scanned}
	
	\acro{2D}{two-dimensional}
	\acro{3D}{three-dimensional}
	
	\acro{AORI}{Atmosphere and Ocean Research Institute}
	
	\acro{AIC}{Akaike information criterion}
	\acro{DL}{deep learning}
	\acro{FPR}{false positive rate}
	
	\acro{CDP}{common depth point}
	
	\acro{JMA}{Japan Meteorological Agency}
	
	\acro{MCS}{multi-channel seismic}
	
	\acro{KPSDM}{Kirchhoff pre-stack depth migration}
	
	\acro{PSDM}{pre-stack depth migration}
	\acro{RMS}{root mean squared}
		\acro{BSRs}{bottom-simulating reflectors}
	
			\acro{LVZ}{low-velocity zone}
						\acro{HVZ}{high-velocity zone}

	\acro{SRMA}{surface-related multiple attenuation}
	
	\acro{GNSS}{global navigation satellite system}
	
	\acro{RMO}{residual move out}
	
	\acro{JSPJ}{Japan Sea Earthquake and Tsunami Research Project}
	
		\acro{MLIT}{Ministry of Land,
			Infrastructure, Transport and Tourism}

	\acro{AGC}{automatic gain control}
	
	\acro{KT}{Katatang thrust}
	\acro{STDS}{South Tibetan Detachment System}
	\acro{GSI}{Geological Survey of India}
	\acro{GSB}{Geological Survey of Bhutan}
	\acro{AI}{artificial intelligence}
	\acro{ITSZ}{Indus-Tsangpo Suture Zone}
	\acro{GPS}{global navigation system}
	
	\acro{GHS}{Greater Himalayan Sequence}
	\acro{REAL}{Rapid Earthquake Association and Location}
	\acro{IRIS}{Incorporated Research Institutions for Seismology}
	\acro{FNR}{false negative rate}
	\acro{STEAD}{STanford EArthquake Dataset}
	\acro{SNR}{signal-to-noise ratio}
	\acro{$M_L$}{local magnitude}
	\acro{ADLWs}{advanced detection and location workflows}
\end{acronym}
\bibliographystyle{plainnat}
\bibliography{references}
\section*{Acknowledgements }

The authors would like to thank \href{https://www.aspentech.com/}{Aspen Technology Inc.} for providing the seismic data processing software used in this research. We also thank the crew of the \textit{R/V Hakuho-Maru} for their assistance in acquiring the seismic reflection data for this study.

\end{document}